\documentstyle[12pt,epsf,aaspp4]{article}

\newcommand{\myshowfig}[1]{#1}

\newcommand{\comment}[1]{}
\tighten
\def\be{\begin{equation}}
\def\ee{\end{equation}}
\def\myA{{\bf C}^{-1}}

\begin{document}

\title{Magnification Ratio of the Fluctuating Light in
Gravitational Lens 0957+561}

\author{
William H. Press and George B. Rybicki
}
\affil{
Harvard-Smithsonian Center for Astrophysics
}
\authoraddr{
60 Garden Street, Cambridge, MA 02138
}

\begin{abstract}
Radio observations establish the B/A magnification ratio of
gravitational lens 0957+561 at about 0.75.  Yet, for more than 15
years, the {\it optical} magnfication ratio has been between 0.9 and
1.12.  The accepted explanation is microlensing of the optical source.
However, this explanation is mildly discordant with (i) the relative
constancy of the optical ratio, and (ii) recent data indicating
possible non-achromaticity in the ratio.  To study these issues, we
develop a statistical formalism for separately measuring, in a unified
manner, the magnification ratio of the {\it fluctuating} and {\it
constant} parts of the light curve.  Applying the formalism to the
published data of Kundi\'c et al. (1997), we find that the magnification
ratios of fluctuating parts in both the g and r colors agrees with
the magnification ratio of the constant part in g-band, and
tends to disagree with the r-band value.  One explanation could
be about $0.1$ mag of consistently
unsubtracted r light from the lensing galaxy G1, which seems unlikely.
Another could be that 0957+561 is approaching a caustic in the
microlensing pattern.
\end{abstract}


\section{Introduction}\label{sect:intro}

The gravitational lens system 0957+561 has by now been observed at
optical and radio wavelengths for nearly twenty years (Walsh,
Carswell, and Weymann 1979; Porcas et al. 1979).  Radio studies have
definitively established that the B/A magnification ratio of the lens,
measured at the core of the radio images (which lies at the location
of the optical point source images), is close to 0.75; some recent
measurements are $0.75\pm 0.02$ (Garrett et al. 1994), $0.752\pm
0.028$ (Conner et al., 1992).  Note that while there is some
controversy about the radio magnification ratio at the location of the
radio jet, as opposed to the core (see, e.g., Garrett et al. 1994),
only the core value interests us here.

Because the variability of the quasar in the optical is larger than in
the radio, measurement of the B/A magnification ratio in the optical
requires that the light curves be shifted by the correct time delay
$\tau$ before the ratio is taken.  Thus, the earliest determinations
of B/A were incorrect.  For example, Young et al. (1980) obtained
a ratio of 0.76, comfortingly -- yet erroneously -- close to the
radio magnification.

However, at least from Vanderriest et al. (1989) on, who used a value
for $\tau$ quite close to definitive recent determinations (Kundi\'c et
al. 1997), it has been clear that the B/A magnification ratio of the
optical continuum in the B and A point sources is quite different from
0.75, and moreover has remained at least fairly constant for the full
history of observation.  Smoothing over observing
seasons, Vanderriest et al. (1989) obtained a B/A ratio varying between about
0.9 and 1.05 over the observing seasons early-1980 through early-1986
(times referenced to A component), with a single best-fit value of 0.97.
It is debatable whether the variation around the best-fit value is actual
time variation of the lens magnification ratio (as distinct from
time variation in the quasar luminosity, n.b.) or observational artifact.
However, it does seem quite likely that the magnification ratio varied
by no more than about $\pm 8$\% during this time.

More recently, the value recently obtained by Kundi\'c et al. (1995, 1997) for
the 1995 season (A component) is $1.12\pm 0.01$ in g-band (with the
error bar, a 95\% confidence limit, depending somewhat on the method
of reduction used).  So, it is quite plausible (and not contradicted
by other measurements in the literature) that the optical B/A remained
in the range 0.9 to 1.12 from 1980 through 1995, and possible that the
variation has been considerably smaller than this range.

The discrepancy between the optical and radio magnification ratios has
long been understood as due to microlensing (as predicted by Chang and
Refsdal 1979, and Gott 1981).  The proper radius of the Einstein ring
from a 0.5 $M_\odot$ star at the lens galaxy redshift $z=0.36$,
illuminated by the quasar at redshift $z=1.41$, is about $2\times
10^{16}\,h^{-1/2}$cm.  Since the radio emission region is much larger
than this scale, it averages spatially over the microlensing pattern
and is magnified by the macrolens ratio of 0.75.  If the optical
magnification indeed differs by $\sim 30$\% from the macrolens value,
then the optically emitting region must be smaller, or at most a few
times larger, than the Einstein ring scale.  This accords nicely with
(e.g.)  the size of an accretion disk smaller than 100 Schwarzschild
radii around a $10^9$M$_\odot$ black hole (a scale of $3\times
10^{16}$cm).

This Einstein ring radius
is only marginally, however, in accord with the apparent constancy
of the microlensed magnification ratio:  Since the Earth, the microlensing
star (or stars, the effect being collective), and the quasar each have
(3-dimensional) peculiar velocities of at least 300 km/s, the Earth should
move through $\sim 100$\% microlensing variations in $\sim 10$ yr
(see Kochanek, Kolatt, and Bartelmann, 1996 for related calculations).
So, the observed microlensing is about a factor 10 {\it too constant},
and one is invited to speculate on whether something other than
luck is the reason.

Another invitation to speculation is the fact that Kundi\'c et al.
obtain rather different magnification ratios in their r- and g-band
data, with the r ratio being $1.22\pm 0.02$, with, again, the error
bar depending on the method of analysis used.  By any interpretation
of the error bars, however, the r and g results are strongly
discrepant.  (Again note that there is no assumption that the
fluctuations themselves have the same amplitude in the two colors,
but only that the magnification {\it ratio} should be the same.)
Either a full $0.1$ mag of r-band galaxy light has
escaped Kundi\'c et al.'s careful subtraction in the B image, or
something else is going on in the lens magnification ratio.

With these two hovering peculiarities (possible excess time-constancy,
and possible non-achromaticity), it seems useful to try to get additional
information on the magnification ratio.  This paper therefore asks the
questions: Is the optical magnification ratio the same for the source region
that produces the {\it fluctuations} in quasar light as it is for the
source region that produces the {\it constant} light?  And, does the
magnification ratio of the fluctuations (which we may call the ``AC''
magnification ratio or ``ACMR'') agree more closely with
the r- or g-band magnification ratio previously measured (here called
the ``DC'' magnification ratio or ``DCMR'')?

The answers to these questions can help diagnose the following
situations: (i) If, as is true in many models, the size of the
emitting region is much smaller than the microlensing scale,
then all the magnification ratios should have the same value.
(ii) If there is a problem with r-band galaxy subtraction -- or
any other constant source of flux added to one lens component and not
the other -- then the ACMR should represent the ``true'' microlens
magnification ratio, and we might further expect it to be close to the
g-band DCMR (where galaxy subtraction is a much smaller
effect).  (iii) If the optically emitting quasar accretion disk has a
scale comparable to the microlensing scale, and has (as seems
almost inevitable) color gradients, then the r and g ACMRs, and
r and g DCMRs, might all be distinct.  Indeed, the two ACMRs and two DCMRs
then provide four distinct windows on the convolution of the
accretion disk source with the microlensing pattern.

Conceptually, one measures a DCMR and an ACMR as follows:
Shift one of the light curves (A,B) in time by $\tau$ to undo the lens
delay.  Fit each light curve by a constant value plus a residual
time-varying part.  The ratio of the constant values is the DCMR.
Now, for the two time-varying residuals, fit for a model that makes
the B residual a constant times the A residual.  The best-fitting
constant is the ACMR.

This conceptual formulation, while simple, is actually not quite
right.  In the next section, we will give a statistical formulation of
the problem that is more complete, and also more directly applicable to
unevenly sampled data.  In Section 3, we discuss some
implementation details, and in Section 4 we apply the formulation to the
published data of Kundi\'c et al. (1995, 1997).  Section 5 is discussion and
conclusions.

\section{Derivation of Statistical Method}\label{sect:method}

The spirit of this derivation is very much the same as described in
Rybicki and Press (1992, hereafter RP92), which the reader might wish
to consult at this point.  For the purposes of calculation, we assume
that the underlying fluctuating light curve is generated by a Gaussian
process $f(t)$.  This is of course only an approximation, and we will
comment in Section 3, below, on its limits of validity; but the
Gaussian assumption provides a clean analytic framework.
Moreover, it is actually quite hard to find evidence for non-Gaussianity
in the 0957+561 light curve (see Press and Rybicki, 1997).

The process $f(t)$ is completely characterized by its covariance
\be S(t_1,t_2) \equiv \left<f(t_1)f(t_2)\right> \ee
so that the $\chi^2$ of a series of (for now, noiseless) observations
$f(t_1), f(t_2), \ldots , f(t_N)$ is
\be \chi^2 = {\bf f}^T {\bf S}^{-1} {\bf f} \ee
where ${\bf f}$ is the vector of $f$ values and ${\bf S}$ is
the matrix of $S$ values relating all pairs of times occuring in
${\bf f}$.  The relative probability of a given sequence ${\bf f}$
occuring is
\be P({\bf f}) \propto (\hbox{det }{\bf S})^{-1/2}
  \exp \left(-{1\over 2}\chi^2\right) \ee

We now model the observed A- and B-component light curves as
\begin{eqnarray}
  a(t) &=& a_0 + f(t) + n_a(t) \nonumber \\
  b(t) &=& b_0 + Rf(t) + n_b(t)
\end{eqnarray}
Here $f(t)$ is a Gaussian process, as above, $n_a$ and $n_b$ are the
(assumed Gaussian) noise processes in the A and B measurements,
respectively, and $R$ is the desired B/A magnification ratio, the
ACMR.  (The DCMR would be $b_0/a_0$.)

Evidently $a(t)-a_0$ and $b(t)-b_0$ are Gaussian processes.  If ${\bf
a}$ is a vector of A measurements, ${\bf b}$ is a vector of B
measurements, then the $\chi^2$ value of a vector combining both
measurements is immediately
\be \chi^2 =
\left( \begin{array}{c} {\bf a}-a_0 \\ {\bf b}-b_0 \end{array} \right)^T
{\bf M}^{-1}
\left( \begin{array}{c} {\bf a}-a_0 \\ {\bf b}-b_0 \end{array} \right)
\ee
where the matrix ${\bf M}$ has the block-partitioned form
\be {\bf M} =
\left( \begin{array}{cc} {\bf S}_{aa}+{\bf N}_a & R {\bf S}_{ab} \\
 R {\bf S}_{ba} & R^2 {\bf S}_{bb}+{\bf N}_b \end{array} \right) \ee
with ${\bf N}_a$ and ${\bf N}_b$ the noise covariances (here and subsequently
assumed to be diagonal), and the notation ${\bf S}_{ab}$ (e.g.) indicating
a matrix of covariance values $S$ relating the times of A observations
and the times of B observations.  The associated Gaussian probability,
where we now emphasize the dependence on the unknown parameters, is
\be P({\bf a},{\bf b} | a_0,b_0,R) \propto 
(\hbox{det }{\bf M})^{-1/2}
  \exp \left(-{1\over 2}\chi^2\right) \label{eq:seven}\ee

Noting that
\be {\bf M} = \left(\begin{array}{cc} 1 & 0 \\ 0 & R\end{array}\right)
{\bf C} \left(\begin{array}{cc} 1 & 0 \\ 0 & R\end{array}\right) \ee
where
\be {\bf C} = \left( \begin{array}{cc} {\bf S}_{aa}+{\bf N}_a & {\bf
 S}_{ab} \\ {\bf S}_{ba} & {\bf S}_{bb}+{\bf N}_b \end{array}
 \right) \ee
we also have
\be \chi^2 =
\left( \begin{array}{c} {\bf a}-a_0 \\ R^{-1}{\bf b}-R^{-1}b_0
\end{array} \right)^T {\bf C}^{-1}
\left( \begin{array}{c} {\bf a}-a_0 \\ R^{-1}{\bf b}-R^{-1}b_0
\end{array} \right)
\label{eq:ten}\ee
and
\be \hbox{det }{\bf M} = R^{2M} \hbox{det }{\bf C} \label{eq:tenprime}\ee
where $M$ is the length of the vector ${\bf b}$ (the number of B-component
data points).

We now arrive at a somewhat subtle, though important, issue: We do
{\it not} want to assume that the fluctuating process $f(t)$ has zero
mean.  Indeed, physically, $f(t)$ might be everywhere positive, since
a process can emit positive photons but not negative ones.  Thus, in
the context of equation (\ref{eq:ten}), we want to use an unbiased
$\chi^2$ that is independent of moving some constant flux from $f(t)$
and into $a_0$ and $b_0$ (in correct proportion).  RP92 showed how
such a ``Gauss-Markov'', unbiased process is obtained by adding a term
$\lambda{\bf E}{\bf E}^T$ to ${\bf C}$ and then letting
$\lambda\rightarrow \infty$ before calculating ${\bf C}^{-1}$ (see RP92
for explanation of this notation).  Henceforth, we assume that this
limit is always taken.  In this case, $\chi^2$ becomes independent
of any constant added to the vectors in the quadratic form, and
equation (\ref{eq:ten}) can be written as
\be \chi^2 =
\left( \begin{array}{c} {\bf a} \\ R^{-1}{\bf b}-c_0
\end{array} \right)^T {\bf C}^{-1}
\left( \begin{array}{c} {\bf a} \\ R^{-1}{\bf b}-c_0
\end{array} \right)
\label{eq:eleven}\ee
where
\be c_0 \equiv R^{-1}b_0-a_0 \ee

The point is that, for fluctuations {\it not} known to have
zero mean, the three parameters $R$, $a_0$, and $b_0$, are
not, even in principle, separately measurable.  But the two
parameters $R$ and $c_0$ {\it are} measurable.  We refer to
$c_0$ as the ``contamination'', since in the ideal case
that B alone is contaminated by spurious constant light
(for example, an unsubtracted component of the lens galaxy's light)
then $Rc_0$ represents the spurious flux.

It is possible to separate, analytically, the contribution to
$\chi^2$ of $c_0$ and $R$, as follows:  Let
\be {\bf y} \equiv \left( \begin{array}{c} {\bf a} \\ R^{-1}{\bf b}
\end{array} \right), \qquad {\bf q} \equiv \left( \begin{array}{c} {\bf 0}
\\ {\bf E}_b \end{array} \right) \equiv (0,0,\ldots,0,1,1,\ldots,1)^T \ee
so that equation (\ref{eq:eleven}) becomes
\be \chi^2 = ({\bf y} - c_0{\bf q})^T \myA ({\bf y} - c_0{\bf q})
\label{eq:twelve}\ee
Straightforward matrix ``completion of the squares'' shows that
equation (\ref{eq:twelve}) is the same as
\be \chi^2 = ({\bf q}^T\myA {\bf q})
\left( c_0 - {{\bf q}^T\myA {\bf y} \over {\bf q}^T\myA {\bf q}} \right)^2
+ {\bf y}^T {\bf H} {\bf y} \label{eq:thirteen}\ee
where
\be {\bf H} \equiv \myA - {\myA{\bf q}{\bf q}^T\myA \over
  {\bf q}^T\myA {\bf q}} \label{eq:fourteen}\ee
The probability associated with equation (\ref{eq:thirteen}) is
(using equations \ref{eq:seven} and \ref{eq:tenprime})
\be P({\bf a},{\bf b} | R,c_0) \propto
(R^{2M}\hbox{det }{\bf C})^{-1/2}
  \exp \left(-{1\over 2}\chi^2\right) \label{eq:fifteen}\ee

Equations (\ref{eq:thirteen}) -- (\ref{eq:fifteen}) could be used
directly on data, to obtain (e.g.) maximum likelihood estimates of
$c_0$, which appears explicitly and only in one term,
and $R$, which appears implicitly
only through ${\bf y}$.  However, we want to take the more Bayesian
viewpoint that $c_0$ is a nuisance variable that ought to be
integrated over, rather than maximized.  (In practice, we find
that $c_0$ is generally so well determined at fixed $R$
that it hardly matters whether we are Bayesians or not.)
Bayes Theorem now gives the simple result
\be P(R | {\bf a}, {\bf b}) \propto
\int_{-\infty}^{\infty} dc_0\, P({\bf a}, {\bf b} | R, c_0)
\propto R^{-M} \exp \left( - \frac{1}{2} {\bf y}^T {\bf H}
{\bf y} \right) \qquad \hbox{[ACMR]}\label{eq:sixteen}\ee
where, in the last equality, all constant factors that do not
depend on $R$ have been dropped.  The constant of proportionality
is determined, in Bayesian manner, by requiring that the integral
of equation (\ref{eq:fifteen}) over $R$ be unity.
Equation (\ref{eq:fifteen}) is easily evaluated on a given data
set ${\bf a}, {\bf b}$ for a range of values of $R$, giving either
Bayesians probabilities or confidence intervals for $R$, which
is the AC magnification ratio (ACMR, see Introduction).

What about the DC magnification ratio (DCMR)?  From the above
discussion, we now see that it is not directly measurable, without
further assumption.  The reason (to reiterate) is that some part of
the light that is physically part of the fluctuating part, and might
fluctuate in the future, may happen to be constant, or nearly so, over
a finite data set.  The closest thing to what the observer means by
DCMR is obtained by setting $c_0$ to zero in equation (\ref{eq:twelve})
and following equations.  That is, the observer pre-corrects the
data sets ${\bf a}, {\bf b}$ for all known constant sources of
error (galaxy light subtraction, and so on), then fits for the
best single ratio $R$ that directly relates the A and B data sets.  The
Bayesian probability corresponding to equation (\ref{eq:sixteen})
is thus evidently, by equation (\ref{eq:twelve}),
\be P(R | {\bf a}, {\bf b}) \propto
R^{-M} \exp \left( - \frac{1}{2} {\bf y}^T \myA
{\bf y} \right) \qquad \hbox{[DCMR]}\label{eq:seventeen}\ee
We will see that this DCMR is generally much better determined
statistically than is the ACMR, but at the price of having unknown
systematics (in the pre-correction of the data).  The ACMR is
much less well-determined, but is completely independent of such
systematics.  Thus, there is a synergy in computing both
magnification ratios by the unified formalism given here.

\section{Hints and Limitations}

We need to discuss how limiting is the original assumption of
a Gaussian process.  The Gaussian assumption enters in two ways:
First, it provides an ``automatic'' means of interpolating
across the time intervals between measured A and B data points (which
don't in general line up after shifting one by $\tau$).
Experience has shown (also see Press, Rybicki, and Hewitt 1992a,b)
that this use is quite robust -- the interpolation is sensible
and not very different from any other sensible method.

Second, the Gaussian assumption is used in associating $\chi^2$
values (and, more importantly  $\Delta\chi^2$ values) to
probabilities.  One should definitely be suspicious of this
association in the tails of the distribution.  If, however, one
takes the resulting probability distributions as indicative of
central value and uncertainty, rather than as correct in detail,
then one is on relatively safer ground:  The procedures described
are essentially $\chi^2$ parameter estimations, and such estimations
(in the limit of large numbers of data points, when the
central limit theorem can apply) do not require any additional Gaussian
assumptions.

As in any $\chi^2$ parameter estimation, the method is valid only if
the error model is close to correct.  A simple test to be passed is
whether the values of $\chi^2$ in the exponentials of equations
(\ref{eq:sixteen}) and (\ref{eq:seventeen}) are consistent with the
number of data points, i.e., whether the reduced $\chi^2$ is close to
unity.  When this is the case, we have found that additional fine
tunings of the error model (fiddling the functional form of the
covariance in the matrix ${\bf S}$, or tradeoffs between adjusting the
correlation model embodied in ${\bf S}$ and the noise model embodied
in ${\bf N}$) have little effect on the output $P(R)$ probability
functions.  If, however, the original reduced $\chi^2$ is not close to
unity (whether too low or too high), then any kind of rescaling
procedure must be viewed as introducing unknown biases into the method
given here.

Indeed, the reason that we restrict ourselves, in the next Section, to
the data of Kundi\'c et al. (1995, 1997) is that for other data sets (e.g.,
Vanderriest et al. 1989) we have not been able to formulate a satisfactory
and self-consistent error model (${\bf S}$ and ${\bf N}$ matrices).

A final technical note is to point out that the calculations
implied by equations (\ref{eq:sixteen}) and (\ref{eq:seventeen})
can all be done using the ``fast methods'' for inverting
${\bf S}+{\bf N}$ matrices described by Rybicki and Press (1995).
These fast methods restrict the correlation model to a particular
functional form, basically requiring the correlation
structure function to be a linear function of lag time.  However,
this approximation appears to be adequate for the 0957+561 data,
at least for these purposes, and the resulting reduction in
computation time is enormous, and would become essential for very large
data sets.

\section{Application to Kundi\'c et al. Data}\label{sect:apply}

Figure 1 shows the results of applying the analysis described in the
previous two sections to the data set published by Kundi\'c et
al. (1995, 1997).  First we convert the data and error bars from magnitudes
to fluxes $a(t)$ and $b(t)$.  Next,
we use the A component light curves
(separately in the g and r colors) to estimate a structure
function $V(\tau) \equiv \left< [a(t)-a(t+\tau)]^2 \right>$.
Next, we use this structure function, along with the
errors, to construct the ${\bf S}$ and ${\bf N}$ matrices
appropriate for the ``fast'' method (Rybicki and Press, 1995).
It is at this stage that the Gauss-Markov (unbiased) limit is
taken (see discussion before equation \ref{eq:eleven}, above).

The upper panel of the figure plots $P(R)$ as a function of $R$ for
both the DCMR (equation \ref{eq:seventeen}; plotted as the ``narrow''
distributions) and the ACMR (equation \ref{eq:sixteen}; plotted as
the ``broad'' distributions).  In each case, the results for the
g-band data are shown as solid curves, while the r-band data is shown
as dotted curves.

Our DCMR values closely reproduce the B/A flux ratios quoted by Kundi\'c
et al. in both the r and g bands, and our errors (widths of curves
shown) are comparable to the latter's quoted errors.  Our ACMR values
-- novel to this work -- are seen to be compatible, in both g and r,
with the g-band DCMR.  Although there is no strict incompatibility
among all four values (except the two DCMRs, as previously remarked)
the distributions in Figure 1 tend to support the conclusion that the
r-band DCMR is ``odd man out''.

The lower panel in Figure 1 shows how many magnitudes of contamination
(relative to the time-average flux) would need to be present in the B
image to move the DCMR peak from its plotted location to any other
location.  By definition, the curves go through zero at the DCMR peak
centers.  One sees that about $0.1$ mag is required in r-band to shift
the DCMR to compatibility with g-band; however, even without shifting,
the r-band DCMR is at least marginally compatible with the r-band
ACMR.

\section{Discussion and Conclusions}

While these data, in this analysis, do not support any very definitive
conclusions, we may make the following remarks:

Occam's razor would seem to indicate that the r-band light curve of
Kundi\'c et al. has about $0.1$ mag of residual, unsubtracted, constant
light, as perhaps from unmodeled small-scale variations in the lens
galaxy surface brightness.  If this is the case, then all the data are
compatible with a single magnification ratio for both colors and for
both the fluctuating and constant pieces.  This in turn suggests an
accretion disk scale much smaller than the microlensing scale, in
accord with theoretical prejudice.  The utility of the ACMRs is that,
taken together, they strongly favor the hypothesis that the g-band
magnification ratio is the correct one, and that nothing more exotic
is going wrong.  We note, however (per E. Turner, private
communication), that the galaxy G1 is something like 2 magnitudes
fainter than component B in r band; thus the amount of unsubtracted light
would need to be comparable to the total brightness of G1, which
seems quite unlikely.

It is up to the observers, not us, to decide whether Occam's razor
should rule in this case.  If $0.1$ mag of residual is not possibly
present, then we must conclude that the accretion disk scale is
comparable to the microlensing scale, and that the constant r-band
part of the disk is {\it more} strongly magnified than (at least some
of) the other three regions.  It seems likely on physical grounds that
the fluctuating regions should be smaller than the constant regions,
and that the g-band regions should be smaller than the r-band regions
(temperature decreasing outward in the disk).  For the larger (r-band
and constant) region to have a higher magnification ratio than an
enclosed smaller region, the larger region must extend to a place
where the magnification is a superlinear function of position in the
sky.  This might indicate at least a fair chance of the B image
passing through a caustic in the near ($\sim 10$ year) future.  This
possibility, as well as the reconciliation of the relative constancy
of the magnification ratio over the last 15 years, will be explored by
Monte Carlo simulations in another paper (Press and Kochanek, in
preparation).

\acknowledgements

This work was supported in part by NSF grant PHY-9507695 at Harvard,
and PHY-9513835 at the Institute for Advanced Study, whose hospitality
is acknowledged.  We thank Chris Kochanek, Ed Turner, Ramesh Narayan,
and Paul Schechter for helpful discussions.

\newpage

\begin{figure}[h]

\centerline{\sc Figure Captions}
\myshowfig{\epsfxsize=6.0in\epsfbox{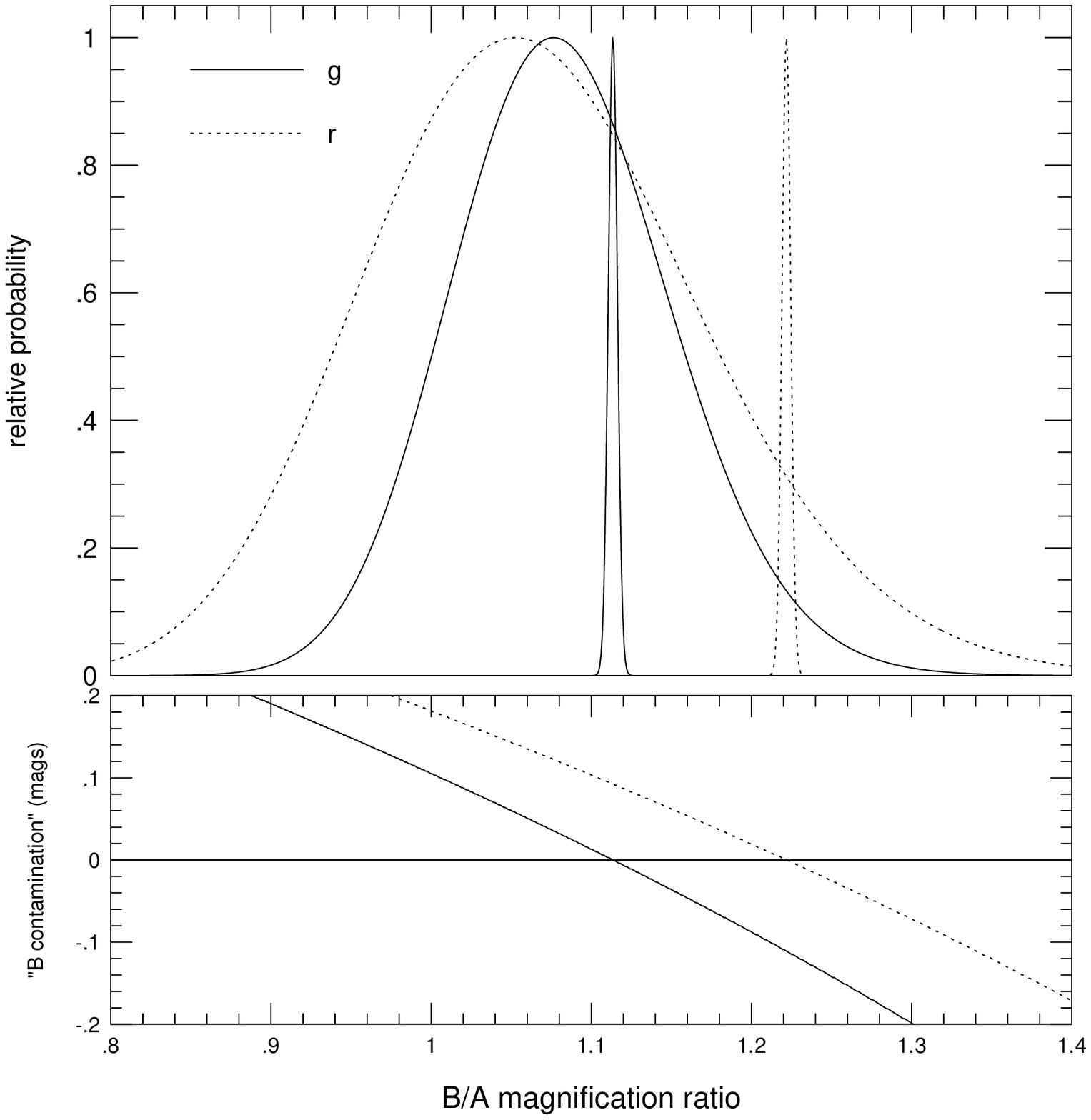}}
\caption{Results of applying the formalism of this paper to the
0957+561 data of Kundi\'c et al. (1997).  The upper panel shows Bayesian
probability distributions for the B/A magnification ratios in two
colors (solid versus dotted curves) and by two analysis techniques.
The narrow distributions reproduce traditional analyses that look for
a single flux ratio relating the A and B light curves.  These give
statistically narrow results (as shown), but have unknown systematics
associated with (e.g.) the background subtraction of light from the
lensing galaxy, especially the B component in r color.  The broad
distributions, new in this work, are the magnitude ratios obtained by
using only the fluctuating part of the measured light curves.  While
statistically less accurate, these are completely insensitive to
background subtraction.  The lower panel shows, for each color, the
amount of constant light necessary to subtract from the B image to
move the narrow distributions to different magnification ratios.
See text for further interpretation of these results.}
\label{fig:espec}
\end{figure}

\end{document}